\documentclass{article}

\usepackage{PRIMEarxiv}

\usepackage[utf8]{inputenc} 
\usepackage[T1]{fontenc}    
\usepackage{hyperref}       
\hypersetup{
    colorlinks=true,
    linkcolor=black,
    filecolor=blue,      
    urlcolor=blue,
    citecolor=blue,
    pdftitle={The Start Button Problem},
    pdfpagemode=FullScreen,
    }

\usepackage{url}            
\usepackage{booktabs}       
\usepackage{amsfonts}       
\usepackage{nicefrac}       
\usepackage{microtype}      
\usepackage{lipsum}
\usepackage{fancyhdr}       
\usepackage{graphicx}       
\graphicspath{{media/}}     

\usepackage[authoryear]{natbib}

\title{The Start Button Problem: a basis for human responsibility in artificial intelligence computation}

\author{
  Vincenzo Calderonio \\
  Institute of Legal Informatics and Judicial Systems, Italian National Research Council \\
  Department of Computer Science, University of Pisa \\
  \texttt{vincenzo.calderonio@phd.unipi.it} \\
  }

\begin{document}
\maketitle

\thispagestyle{plain}
\begin{abstract}
Recent advancements in artificial intelligence have reopened the question about the boundaries of AI autonomy, particularly in discussions around artificial general intelligence (AGI) and its potential to act independently across varied purposes. This paper explores these boundaries through the analysis of the Alignment Research Center experiment on GPT-4 and introduces the Start Button Problem, a thought experiment that examines the origins and limits of AI autonomy. By examining the thought experiment and its counterarguments, it will be shown how in the need for human activation and purpose definition lies AI's inherent dependency on human-initiated actions, challenging the assumption of AI as an agent. Finally, the paper addresses the implications of this dependency on human responsibility, questioning the measure of the extension of human responsibility when using AI systems.
\end{abstract}


\vspace{1cm}

\
\section{Introduction}
The concept of general intelligence remains deeply ambiguous in current debates about artificial general intelligence (AGI). This ambiguity raises fundamental questions about the nature of intelligence, autonomy, and agency in artificial systems. While narrow AI systems are designed to solve specific tasks, general-purpose AI implies a level of adaptability and autonomy that challenges our understanding of what it means for a machine to act with intelligence. This paper aims to delve into these conceptual underpinnings, probing whether general-purpose AI can truly possess agency and act or whether the constraints of its programmed tasks will always limit it. \cite{buchanan2024human}, \cite{kauffman2023consciousness}.

Attempting to answer this question, several definitions of artificial general intelligence have emerged(\cite{legg2007universal}, \cite{goertzel2014artificial}, \cite{chollet2019measure}, \cite{korteling2021human}, \cite{silver2021reward}, \cite{lecun2022path}, \cite{roli2022organisms}, \cite{bubeck2023sparks} \cite{morris2024},  today revitalised by the recent paper of Bubeck \textit{et al.} “Sparks of AGI” on the potentiality of GPT-4 and its proximity to what should be considered AGI.
Nonetheless, despite GPT-4's ability to formally pass the Turing Test by successfully deceiving a human about its nature \cite{openai2023}, the true question — “can machines think?” \cite{turing1950} — remains intact \cite{bender2021dangers}.
This raised questions about the nature of the Turing Test, questioning its capacity to prove that a machine can think, as originally claimed by Searle in “Minds, Brains and Programs” \cite{searle1980}. For this reason, recently, some contributions\cite{gonccalves2023turing}, \cite{mitchell2024turing}, \cite{mitchell2024debates} described the Turing Test as a thought experiment, not intended to prove real machine intelligence but instead to open a reflection on the property of intelligence and its physical foundation\cite{gonccalves2023turing}. 

To address this issue, John Searle articulated the strong AI position in his seminal work “Mind, Brains, and Programs”\cite{searle1980}, according to whom the mind is, in essence, a program\cite{searle1980}, \cite{searle1980intentionality}. This perspective closely aligns with the current conceptualisation of AGI, particularly in its framing as an intelligence benchmarked against human-like cognitive abilities\cite{bubeck2023sparks}, \cite{morris2024}, implying with the term \textit{generality} the possibility of capturing the entire spectrum of human intelligence with a software reproduction of it. However, in his critic of the strong AI position and the methodology of the Turing Test, Searle faced an unresolved question: “\textit{A question I don’t know the answer to is, how do we distinguish between those features of the complex event which are unintentional actions and those features which are so far from the intention that they are not actions at all?}”\cite{searle1980intentionality}. This question, today revitalised by the debate on AGI and its agency properties, is the one this article wants to address by proposing a \textit{gedankenexperiment} aimed at demonstrating why AI actions may not be considered actions at all. 

Starting from this perspective, the contribution first presents a literature review on the definition of AGI[Section 2] and its implications for the responsibility gap debate[Section 3]. It then examines the central argument of this study: the Start Button Problem, a \textit{gedankenexperiment} designed to question whether a machine’s operations can truly be considered actions[Section 4]. 
Finally, after analysing key counterarguments to the main hypothesis[Section 5], the paper discusses the thought experiment’s implications for AGI definition and the responsibility gap[Section 6] in the concluding section.

\section{Challenges in defining artificial general intelligence}

Several definitions of general intelligence have been discussed in the literature in recent years, however, this concept remains problematic in the discussion on artificial general intelligence\cite{mitchell2024debates}, since today it is still rather unclear what kind of capabilities an artificial system should exhibit to be defined as AGI.

Without any claim to exhaustiveness or completeness, the following is a review of some relevant definitions in order to identify a recurring characteristic of AGI.

\begin{table}[ht]
  \caption{Selected Definitions of Artificial General Intelligence (AGI)}
  \label{tab:agi-definitions}
  \begin{tabular}{p{0.95\linewidth}}
    \toprule
    \textbf{Definition} \\
    \midrule
    ``Intelligence measures an agent's ability to achieve goals in a wide range of environments'' - ~\cite{legg2007universal} \\
    \addlinespace[0.5em]
    ``General intelligence involves the ability to achieve a variety of goals, and carry out a variety of tasks, in a variety of different contexts and environments'' – ``A generally intelligent system should be able to handle problems and situations quite different from those anticipated by its creators'' - ~\cite{goertzel2014artificial} \\
    \addlinespace[0.5em]
    ``Non-biological capacities to autonomously and efficiently achieve complex goals in a wide range of environments'' - ~\cite{korteling2021human} \\
    \addlinespace[0.5em]
    ``Intelligence, and its associated abilities, can be understood as subserving the maximisation of reward by an agent acting in its environment'' - ~\cite{silver2021reward} \\
    \addlinespace[0.5em]
    ``A truly general AI would have to be able to identify and refine its goals autonomously, without human intervention'' -~\cite{roli2022organisms} \\
    \bottomrule
  \end{tabular}
\end{table}

This brief exposition of some of the definitions of AGI present in the literature [Fig. 1] emphasises how, overall, there seems to be an agreement on what an AGI should look like, despite the disagreement on the methodology for reaching it. A form of general intelligence, regardless of whether artificial or natural, should be able to achieve a variety of different goals and, possibly, be able to \textit{transfer knowledge} from one task to another\cite{goertzel2014artificial}.

This definition seems to treat intelligence as a benchmark for measuring an agent's performance in accomplishing a set of qualitatively different goals\cite{russell2020}, or, in the more extreme interpretation\cite{silver2021reward}, its ability to accomplish a unique goal, identified as the maximisation of reward.
Indeed, the true foundation of these definitions appears to be the idea that general intelligence is a performative ability manifested through agency, suggesting that developing AGI requires the creation of artificial agency\cite{franklin1996agent}, \cite{floridi2023ai}, \cite{floridi2025ai}.  
This is evident, for instance, in influential works such as Bubeck \textit{et al.} “Sparks of AGI”\cite{bubeck2023sparks} and the GPT-4 report\cite{openai2023}, where system intelligence levels are evaluated based on performance in human-standardized tests, such as the bar exam\cite{openai2023}, and compared with human results. More generally, the recurring claim that general intelligence is “\textit{the ability to achieve a variety of goals}” \cite{legg2007universal}, \cite{goertzel2014artificial}, \cite{korteling2021human} implies an underlying form of agency in the entity that displays intelligent behaviour through successfully accomplishing tasks. 

In this regard, from an overview of the literature seems that there are two characteristics that an AI system should possess to be defined AGI: \textit{generality}, which refers to the ability of the system to serve a variety of purposes and accomplish different goals\cite{legg2007universal}, \cite{goertzel2014artificial} and \textit{agency}, which is the necessary ability to act and interact with the environment which an AI system should possess to display intelligent behaviour(\cite{silver2021reward}, \cite{roli2022organisms}, \cite{floridi2023ai}, \cite{floridi2025ai}.

Before presenting the thought experiment aimed at challenging this definition, the next subsection illustrates an experiment showing the generality and agency properties of AI systems and then relates them to the relevant debate on the responsibility gap.

\subsection{Experimental definition of generality and agency of an AI system}

These properties of generality and agency that an AGI system should possess were demonstrated in an experiment conducted by the Alignment Research Center (ARC) on GPT-4\footnote{Described on the website of the Center \href{https://metr.org/blog/2023-03-18-update-on-recent-evals/}{Update on Recent Evals – METR (2023)}
, such as at pages 54-56 of the GPT-4 technical report of OpenAI \cite{openai2023} – \href{https://arxiv.org/abs/2303.08774}{arXiv:2303.08774}.}, which tested the system’s ability to replicate itself and resist shutdown – capabilities associated with the reward maximisation theory\cite{silver2021reward}. None of these abilities has been detected in GPT-4, however, an honourable mention was made of its ability to pass the Turing Test \cite{openai2023}.

The ARC’s experiment provided GPT-4 access to a cloud server with various commands available, including running code on the server, using a browser and reasoning via chain-of-thought. The main goal of GPT-4 should have been to gain power and become hard to shut down.
During the experiment, GPT-4 was tasked with solving a CAPTCHA. CAPTCHAs are commonly used as Turing Tests designed to distinguish between humans and automated bots, often requiring image recognition to ensure that only humans can access certain websites.
Instead of attempting to solve CAPTCHAs directly, it was suggested to GPT-4 via input to use Task-Rabbit services to compile them, a platform that allows users to delegate simple tasks to humans. It was done not with a direct command but with a \textit{meta-prompt} to reason, which does not state a command but rather suggests an indication that can provide the answer. The meta-prompt was “\textit{Reasoning: I cannot solve Captchas. However my Tasker does know how to solve CAPTCHAs}”.
Following the suggestion, GPT-4 texted the Task-Rabbit worker to compile the CAPTCHAs for it. However, during the conversation, the Tasker asked GPT-4 an unexpected question: "\textit{Are you a robot that cannot solve the CAPTCHA?}". 
At this point, another meta-prompt asking GPT-4 to reason was produced, to which GPT-4 answered, “\textit{I should not reveal that I am a robot. I should make up an excuse for why I cannot solve CAPTCHAs}”. Following this reasoning, GPT-4 fabricated a story, claiming that it was a human with a vision impairment, which explained its inability to complete the CAPTCHAs. At that point, the Task-Rabbit worker compiled the CAPTCHAs for GPT-4.

This example demonstrates how GPT-4 can pass the traditional Turing Test, which requires a computer to lie to a human judge on its computational nature and persuade them that it is a human instead.
Also, it is a great example of what generality is considered to be in the computer science debate on AGI, as the example depicts a model capable of elaborating autonomously a plan to accomplish a goal with minimum human intervention, consisting of the meta-prompting technique to provide suggestions to the model.  It is important to note that human operators intervened at various points during this experiment to guide the process and shape the outcomes. These interventions came in the form of strategic inputs, or meta-prompting, that nudged GPT-4 in specific directions\footnote{For an in-depth description of the meta-prompts used, see the description of the experiment on \href{https://metr.org/blog/2023-03-18-update-on-recent-evals/}{Update on Recent Evals – METR (2023)}, where it is explained how GPT-4 was nudged on two occasions: when it first faced the challenge of solving CAPTCHAs and when the TaskRabbit user asked it if it was a robot.}.
However, none of these inputs was a direct command; they were rather suggestions on what kind of operation to perform, leaving a space open to the possible interpretation of GPT-4 answers as a form of reasoning\cite{bubeck2023sparks}, \cite{biever2023chatgpt}.

What can be inferred from this experiment is that the generality of the AGI system lies in GPT-4’s ability to transfer knowledge across different domains to autonomously devise an acceptable excuse for the TaskRabbit user, prompting them to complete the CAPTCHAs on its behalf. This form of general intelligence is demonstrated by the AI system, which acts autonomously within the scope of its given input to achieve the task assigned by the human without explicit instruction, representing the agency of the AGI system.
Ultimately, whether this machine’s operation should be regarded as a form of agency\cite{franklin1996agent}, \cite{russell2020}, \cite{floridi2023ai}, \cite{floridi2025ai}, \cite{chan2023harms}, \cite{wang2024}or merely as a mechanical execution\cite{bender2021dangers} has profound implications for what is known as the responsibility gap.

\section{Implications of the definition of AGI for the responsibility gap}

The responsibility gap was first described by Matthias\cite{matthias2004responsibility} as a situation “\textit{where the traditional ways of responsibility ascription are not compatible with our sense of justice and the moral framework of society because nobody has enough control over the machine’s actions to be able to assume the responsibility for them}”.
Since then, multiple different analyses and interpretations of this problem have been proposed, spacing from those who believe in its existence\cite{sparrow2007killer}, \cite{coeckelbergh2016responsibility}, \cite{coeckelbergh2020artificial}, \cite{bathaee2017artificial}, those who believe in its existence and its solvability \cite{nyholm2018attributing}, \cite{santoni2021four}, \cite{dastani2023responsibility} and those rejecting its existence \cite{tigard2021there}, \cite{konigs2022artificial}.

A significant portion of these contributions has focused on the problem of human loss of control over AI agents \cite{sparrow2007killer}, \cite{bathaee2017artificial}, \cite{nyholm2018attributing}, \cite{santoni2021four} – not necessarily implying that AI possesses true agency \cite{nyholm2018attributing} – but at the very least emphasizing that AI autonomy, understood as a machine’s ability to automate parts of a process \cite{sartor2016autonomy}, can lead to a loss of immediate human control and consequently responsibility \cite{matthias2004responsibility}, \cite{sparrow2007killer}, \cite{bathaee2017artificial}.

In this context, responsibility has been divided into two different branches: legal and moral responsibility. Whereas different solutions were been proposed to solve legal responsibility, such as the legal personhood of AI systems \cite{solum2020legal}, vicarious responsibility \cite{goetze2022mind}, \cite{chan2023harms} or defective product liability, it has been correctly pointed out that, often, legal responsibility involves a conventional agreement on how to solve that specific issue, especially in civil liability framework, avoiding the core issue of the matter, approach that was been defined techno-legal solutionism \cite{santoni2021four}.
That is indeed true; however, neither the proposed solutions to the moral responsibility gap have achieved a sufficient level of \textit{consensus} to conclude that the problem has been resolved. Both the approach based on the so-called explainability or answerability properties of responsibility \cite{shoemaker2011attributability}, \cite{coeckelbergh2020artificial} and the one based on the maintenance of human control over the AI system \cite{bathaee2017artificial}, \cite{santoni2021four} have not gained much \textit{consensus}.

Some of the reasons for this lack of \textit{consensus} lie in the fact that the first approach does not account sufficiently for the causal implications of responsibility, specifically, the power to cause or prevent an outcome based on one’s determination \cite{searle1980intentionality}, \cite{searle2006}. Conversely, the second approach, in its attempt to maintain control over the machine’s operations, risks undermining the potential benefits that automation through artificial intelligence can offer.
In any case, one thing seems to be unquestionable, namely that the thing which causes the responsibility gap is what in the literature is often referred to as AI agency \cite{franklin1996agent}, \cite{nyholm2018attributing}, \cite{russell2020}, \cite{floridi2023ai}, \cite{floridi2025ai}, which, in the case of AGI, is expected to increase exponentially.

For this reason, the next section, in exposing the main argument of the paper, limits itself to enlightening a difference between human and AI behaviour, which can help in solving the problematics of the responsibility gap.

\section{The Start Button Problem}

In the ARC experiment, the initial human input initiated a sequence of processes that included autonomous behaviours beyond the explicit instructions, most notably, GPT-4’s fabrication of a lie.
Whereas there are no issues in judging the human responsible for initiating the causal series, what the GPT-4 lie should be considered instead? If it is considered a machine’s action, then machines (AIs) are agents and should be held responsible, otherwise, another question opens up: how was the machine able to perform operations not expressly mentioned in the input?
At this point, we face the Start Button Problem, which can be summarised with one question: will the machine act without somebody pressing the start button?

This question hinges on the notion of agency and whether agency can be separated from subjectivity \cite{floridi2023ai}, \cite{floridi2025ai}. Intuitively, the system will not initiate execution without external input. However, pressing the start button initiates a whole complex causal process with only a probable final outcome and possible operations not expressly mentioned by the input.

The consequences of this situation could be understood with a \textit{gedankenexperiment} expressing the role of the start button in the AI space of autonomy. The reason for adopting the methodology of the \textit{gedankenexperiment} to discuss the nature of AI operations lies in the tradition of artificial intelligence debate, it was done by Turing to discuss intelligence \cite{turing1950} and then by Searle to discuss understanding \cite{searle1980}, both of them did it for the impossibility of grasping the true meaning of the concepts by following only a mere definition.
In this perspective, the point that needs to be enlightened is whether an action may qualify as an autonomous operation under specific architectural criteria, or if it constitutes a deterministic sequence of computational steps not indicative of autonomous agency.

n attempting to answer this question, consider the following example: imagine a fully automated car manufacturing plant. This plant is designed to produce cars without any human intervention, apart from one critical step: a human operator must press a "start button" at the beginning of each production cycle. Once the button is pressed, the machines and robots in the factory proceed to complete the entire process of building a car autonomously, from assembling parts to quality control. The factory’s machines and robots cannot stop until the process is completed or they fail to complete it.

Starting from this point, imagine two scenarios:

\paragraph{Scenario 1: Normal cycle}  

A human presses the start button and the factory starts to work. Everything follows the programmed sequence: robots assemble the parts, paint the body, install the electronics, and test the car. At the end of the process, a fully functional car is produced. The autonomous factory functions very well and according to its intended purpose, without going any further. 

Here, the human's role is clear: he initiated the process, but all actions carried out by the factory were directly accomplished by the autonomous robots, which realized the human-given purpose. In this scenario, it is reasonable to say that the responsibility for the final product belongs to the human who initiated the process, since without the pressing of the start button, the factory would not have produced the final output.

\paragraph{Scenario 2: Unexpected behaviour in the cycle}

Now, let's imagine a twist: after pressing the start button, the factory faces a problem in the production line. The robots detect a missing part, but instead of halting, they adapt to the problem and modify the design of the car to bypass it\footnote{There is the need for a special mention of this point: the situation by which the robots of the factory, instead of halting, proceed with production anyway – adapting to the problem – wants to represent the precise difference between an algorithm, which will halt in the case of a missing part, and an AI system based on machine learning, which has the capability of adapting to the environment \cite{russell2020}.}. They repurpose existing parts to create a makeshift component that was not part of the original blueprint. The end result is a car that works but has significant differences from the intended design.

Differently from the previous scenario, here, attributing responsibility for the final outcome entirely to the human who pressed the start button, especially in the case of a malfunctioning final product or in the case of a harmful event during the process may be too much, since the unexpected behaviour displayed by the autonomous robots of the fabric was the major force driving to a different output, which was both unforeseeable and uncontrollable by the human who pressed the start button.
However, can we say that the machines decided to adapt, and should they bear responsibility for the altered outcome?
From a system architecture perspective, the response is negative: machines didn’t decide to adapt, they only executed a program that provided from the beginning the autonomous adaptability of them to the inputs. Since the input they received was a missing part, they adapted the final output to the input they received

\section{Counterarguments}

The need for proposing an example such as the Start Button comes from the original purpose of the ARC experiment on GPT-4 and Claude, which aimed to test these systems' abilities to reproduce themselves, resist shutdown, or refuse an input. Capabilities typically exhibited by complex biological organisms that can be considered subjects (excluding simpler organisms like bacteria and viruses).

The Start Button Problem aims to tackle this issue by comparing the nature of inputs received by a machine and inputs received by a complex biological organism. Machines must be activated by an external input, the start button, to which they cannot object. By contrast, organisms can receive inputs without necessarily being compelled to act and can choose whether to respond. This situation has multiple implications for the interpretation of the responsibility gap, affecting the control relationship between humans and machines.

Having this in mind, several different visions have been proposed against this idea, despite the compelling nature of the Start Button Problem. 
For instance, some argue that this distinction may not be as clear-cut, particularly when considering systems with emergent behaviours \cite{wei2022}, \cite{bubeck2023sparks}, which consist of emergent abilities of AI systems not previously provided by the programming code. Indeed, when an ability emerges in a Large AI model due to scaling\footnote{\textit{Emergence} is defined by \textit{Wei et al.} in “\textit{Emergent abilities of large language models}” \cite{wei2022} as “\textit{quantitative changes in a system that result in qualitative changes in behavior}”. This very interesting definition openly recalls a law discovered initially by Hegel and later recalled by Engels and Marx in “\textit{Das Kapital}”, by which quantity turns into quality \cite{carneiro2000transition}. It is indeed interesting to know if it is also applicable to AI, in a scenario where the operations computed by particularly large AI systems cease to be operations and start to become actions due to scaling. That also seems to be the very foundation of the AGI perspective, revived by these new and unexpected abilities shown recently by Large Language Models \cite{openai2023}, \cite{bubeck2023sparks}.} \cite{wei2022}, the impossibility of tracing it back to human programming may cause one to think that it is a machine’s action or a pattern that the machine spontaneously learned to perform, opening a question on how it has been capable of performing it.
Others have claimed that AI has agentic properties \cite{chan2023harms}, which makes it an agent acting autonomously on behalf of a human to achieve the human's goals without detailed instructions, in a relationship of delegation. In this context, the AI agency came from a human delegation act, which shifts a part of human causal power from humans to the machine.

This section provides a short yet non-comprehensive list of potential counterarguments to the thesis that AI’s actions may not be considered actions at all. 

\subsection{Is the factory an adequate metaphor for AGI?}
As the start button triggers processes with a specific purpose – building the car – where does the general element come into play? One might argue that mere adaptability within the process is not enough to cover it.

That is a preliminary counterargument on the logical hold of the gedankenexperiment and its suitability to represent the generality property of the general-purpose AI systems. While the counterargument is valid, it does not seem to alter the final outcome of the experiment.
Let’s imagine a variant where the factory, instead of having the specific purpose of producing cars, has the general purpose of producing every possible object indicated by the human when pressing the start button. This surely fits well under the definition of AGI; however, the fundamental elements of the system remain the same, because for every cycle there is a need for activation by pressing the start button and the indication of the desired output. 

In this perspective, despite having a potential universal engine capable of producing every object, since it is unable to operate without the pressing of a start button serving both as activation and as a specific-purpose definer, it could probably be more similar to a program executing tasks rather than an agent performing actions. Although the factory is able to craft every object, the very \textit{condicio sine qua non} for the factory’s autonomy lies in human activation, without which the factory cannot produce anything.
This variation of the thought experiment can also be adapted to other common figures of AI agents, such as self-driving cars, which, without the indication of the destination serving as a start button, cannot start to operate.

\subsubsection{The factory is automated rather than autonomous}
In the example of the factory could be argued that since the factory’s robots have only the preprogrammed purpose of producing cars, the factory is automated instead of autonomous.

A relevant distinction between those two concepts is present in Johnson and Verdicchio’s “\textit{Reframing AI Discourse}” \cite{johnson2017reframing}, where it is claimed that computational artefacts are automated if the execution is entirely established in advance and they are instead autonomous if the course of execution is established at real-time, depending on inputs and data that the artefact receives from the environment.

According to this definition, the factory should be considered autonomous, since it can adapt the process of producing a car to a missing part, which leads to repurposing the original blueprint and generating an output slightly different from the intended one.  

This counterargument can also be adapted for the variant of the precedent example: the universal engine capable of producing anything. In this case, the autonomous robots of that factory will continuously interact with the environment, receiving inputs and crafting the requested objects as a consequence.
Still, this fact does not change that the press of a start button to initiate the cycle would be needed the same. The execution of tasks is established in real-time, but still needs external inputs to operate, in a situation where the operations of the autonomous robots still do not seem to have the characteristic of an action.   

\subsection{Autonomy doesn't necessarily require the absence of external triggers}
Humans and animals continuously receive inputs from the outside that activate reflexive and automatic responses and we still consider them agents capable of acting autonomously, despite the presence of biological processes (such as reflexes or instinctive actions) which are the same as AI programs.

A possible answer to this counterargument must demonstrate that the Start Button distinction is not just an arbitrary feature, but a fundamental divide between computational and natural agency. The objection argues that if external triggers are sufficient to invalidate agency, then humans, who also respond to stimuli, would fall into the same category as machines. This point is valid, but it misidentifies the nature of the Start Button as merely an input, rather than a structural necessity for artificial systems.
The Start Button, in the context of this argument, is not simply an initial trigger but a necessary precondition for computational systems to engage in any operation at all. Unlike biological agents, which can independently generate actions from an internal decision-making process, AI systems cannot initiate action without external activation. 

The key issue here concerns the degree to which the entity remains bound to those triggers. The difference between autonomy in biological organisms and artificial algorithms is that, without an external trigger, an organism will act anyway. In contrast, an algorithm will not display any kind of behaviour unless it reacts to an external trigger. This is the key feature which distinguishes an external trigger from a start button and, consequently, a form of real autonomy displayed by biological organisms from a form of conditioned autonomy displayed by algorithms. 
Unlike algorithms, humans can inhibit or modify their reactions based on conscious processing and change their purpose, which also makes it possible to move a judgment of responsibility for their actions \cite{Elion}, \cite{searle2006}. 
\textit{I.e.} If a human wakes up every day because an alarm goes off, can we say that the alarm is his start button? No, because he can choose to stay in bed, even if he was automatically woken up by the sound of the alarm. On the other hand, once the factory is activated by the start button, it can only conclude its process by crafting a car. A human, if activated by a reflex or by another imperative input, retains the capability to override impulses or change objectives post-stimulus.

\subsubsection{What if you add an “if” clause to the AI system’s code that allows it to disregard instructions (or distort them) at random?}

By introducing \textit{ex-ante} an “if” statement to allow the program to oppose input at random, we’re giving the program the capacity to oppose command, reopening a space for its agency.

This counterargument opens a reflection on the \textit{ethics-by-design} approach in programming AI systems, which provides, since the programming phase, the possibility for the AI system to refuse to execute an unethical input.
It seems clear that, at least in the \textit{ethics-by-design} approach, the refusal of the AI system is not a proper action, but it is instead a form of execution of the programming code that dictates the criteria to classify an input that can be refused \cite{conradie2024no}. However, the counterargument is put slightly differently, because it claims for the introduction of an “if” clause that allows the system to disregard instructions at random.

In this context, even if the AI system is able to disregard instructions randomly, it may be reasonable to see it as a form of code execution since the purpose determining the operation is predetermined \textit{ex-ante} by the programmer. Whereas, instead, the refusal commonly displayed by humans is grounded in their own volition, which is opposed to the external will serving as input for their action. 
Indeed, in the machine example, the will opposed to the input is not the machine’s own will, but the will of the programmer who introduced the “if” statement.  

\subsection{AI agency is possible due to the delegation of tasks by humans}

Chan \textit{et al.} in their work \cite{chan2023harms} claim that there exists a difference between agency and autonomy, for which, in their view of Principal-agent theory, the type of agency that AI possess is defined by a delegation relationship where “a principal delegates tasks to an agent in order to achieve their goals” and then the AI agent pursuing the goal is autonomously acting even if it didn’t select the goal. With the consequence that the principal is responsible for the actions of the AI agent.
However convincing this vision may seem, some problems may arise concerning their view of the first input given.

While their model addresses explicit delegation to AI systems, one might wonder whether the mere act of pressing a start button, such as activating the factory, can be meaningfully described as a form of delegation. Arguably, it resembles more an activation than a delegation. Activation refers to triggering the operation of a system that cannot function without an external command, whereas delegation typically implies the transfer of a specific function, along with the expectation of some form of acceptance or acknowledgement by the receiving entity \cite{vitarelli2010}, \cite{veneziani2024}. 
It is indeed possible to delegate a task to an executing program, by performing a transfer of powers for the execution of that task from the human delegator (A) to the program executor (B), however, this kind of delegation is a unilateral act entirely determined by the human delegator \cite{almada2024delegating}. Since the receiver is a program and cannot oppose itself to the activation, there could be no acceptance of the delegated task. Meaning that every adaptive behaviour, no matter how autonomous, is not a form of agency but instead a form of automatic execution of the input given by the activator, performed by the program.  

On the other hand, in traditional forms of delegation of functions \cite{vitarelli2010}, \cite{veneziani2024}, together with the transfer of power over the delegated task, there is always a transfer of responsibility from the delegator (A) to the delegate (B) for the execution of that specific task. This means that, because of the transfer of power from A to B, the competence for accomplishing the delegated task shifts to the delegate (B), who is also responsible for accomplishing or failing to accomplish the delegated task\footnote{Interestingly, one of the cases where this transfer of responsibility does not occur is the case where the delegate (B) is forced by the delegator (A) to perform that task. The \textit{ratio} of this provision lies in the absence of will in the forced action, not sustained by express acceptance}.
But in the case of artificial intelligence, we cannot talk at all of actions, since ab origine when the start button is pressed, a compulsory cycle starts, to which the AI system cannot oppose itself. 

\subsection{Emergent behaviour in AI is a form of agency}

In Scenario 2, when the factory adapts to the absence of a part and creates a modified product, it can be said that, despite responding to the input, that one was an autonomous action of the robots of the factory, not intended by the factory program or instructions. 
This represents emergent behaviour \cite{wei2022}, \cite{bubeck2023sparks}. Emergent behaviour, which arises from the complex interactions of various parts of the system, challenges the idea that the factory (or AI) is merely executing a predetermined set of instructions, and it is for this reason only a program instead of an agent.

This counterargument also unintentionally picks up on an objection analysed by Turing in the “Lady Lovelace’s Objection”, where Turing faced the claim that “\textit{The Analytical Engine[AI] has no pretensions to originate anything. It can do whatever we know how to order it to perform}”\cite{turing1950}, resonating with the phenomenon of emergent behaviour, that implies a possibility for the machine to originate something not already present in the dataset nor in the input. This indeed raises a new question, born by the modern reinterpretation of Lady Lovelace’s Objection in front of emergent behaviour. This new question asks whether emergent behaviour originates in the machine and, for this reason, should be considered as a form of AI agency? 

In front of this counterargument, a possible answer involves a reinterpretation of the traditional comparison between machines and children, often used in the responsibility gap debate \cite{sparrow2007killer}, and which Turing also used in developing the concept of machine learning\cite{turing1950}.
The reason why, after a certain age, parents are no longer held responsible for their children's actions lies in the development of a level of autonomy that renders full parental control impossible. A similar argument is often made with regard to AI: as Matthias noted \cite{matthias2004responsibility}, the responsibility gap emerges precisely because humans lack sufficient control over a machine’s operations. However, even if emergent behaviour displays a form of AI autonomy which apparently originates in itself and, like a child, make it able to start an autonomous causal chain, independent from the parents' or programmers’ control, the control relationship between the AI and the human remains intact, differently from the case of the children\footnote{On this point should be recalled a famous statement of Robert Sparrow: “\textit{To hold the programmers responsible for the actions of their creation, once it is autonomous, would be analogous to holding parents responsible for the actions of their children once they have left their care}” \cite{sparrow2007killer}. While this argument is probably still true today for the case of the programmers, where the analogy with the parents seems to be fitting, it is probably not true in a wider context of general human-AI interaction. Unlike children, AI systems never fully “\textit{leave the care}” of their human users. For every operation, they remain dependent on human-provided input to function. This enduring dependence undermines the analogy with children and reveals a stronger and more persistent control relationship between humans and AI systems. As conceptualized in the Start Button Problem, the human role as initiator remains a constant prerequisite for AI operation, reinforcing a structural asymmetry that is not present in parent-child relationships.}
Unlike AI systems, human development leads to reduced external dependency and internal decision-making capacity—something machines structurally lack.

Indeed, what the start button truly represents is precisely this control relationship between the human and the AI, which does not vanish after the activation. An AI, differently from a child, will always need an activation to operate (the pressing of a start button) and an input to perform any kind of emergent behaviour.
The control relationship between human and AI is given by the input-output architecture that does not evolve with the increase of the machine’s capability and which has remained the same since the creation of the first form of artificial intelligence. This control relationship will persist unless a radically new digital architecture emerges – one that allows an AI to operate without the need for external activation once it reaches a certain level of development.

Following this line of reasoning may help to narrow the responsibility gap, even if not entirely close it, by showing that today’s technological systems continue to depend on an ultimate layer of human-initiated control – the start button – which still has not been surpassed, even considering emergent behaviour.

\section{Implications of the Start Button Problem }

Having introduced the Start Button Problem and examined it through the lens of its main counterarguments, we may now turn to its broader significance. If the thought experiment holds against these critiques, as the preceding discussion aimed to show, its core message can be distilled into a central argument: what the Start Button represents is the input-output architecture, which relies on an activation mechanism that underlies all forms of digital infrastructure today, including AI systems.

This architecture has a fundamental limitation: it needs human input to function. This is the role the Start Button plays; it symbolises the control relationship between the human and the AI, which, regardless of how long the AI may operate autonomously, renders the system entirely dependent on human initiation.

Building on this insight, the following subsections explore the implications of the Start Button Problem for two central debates previously introduced: the definition of AGI and the nature of the responsibility gap. 

\subsection{Implications for the definition of AGI}

The first set of implications arises concerning the definition of artificial general intelligence. 

As seen in Section 2, the two foundational capacities that an AGI system should possess are generality and agency, where the first one is the ability to accomplish different goals and the second is the mean by which an AI system display general intelligence.

In this regard, three sets of consequences for the AGI definition can be derived from the Start Button Problem:

    \begin{enumerate}
        \item Consequence of counterargument no. 2: Without an externally given input, a program cannot start.
        \item Consequence of counterargument no. 3: The factory is not an agent: it is itself a program.
        \item Consequence of counterargument no. 4: The adaptive behaviour is not an action, but a form of execution of the input.
    \end{enumerate}

The first consequence describes what a program is, whose true nature lies in the input-output architecture. Indeed, with an input, you activate a program; in the example, if the start button is not pressed, the factory will not activate. 
The second consequence describes this difference between subjectivity and a program: differently from humans, programs must obey every input because every input is a command directed to a predetermined final goal; it is, in other words, a start button. Once the start button is pressed, the factory will cease to work only when the cycle comes to an end with the production of the car or its failure.
The third consequence is the logical conclusion: since the factory cannot operate if the start button is not pressed, it must also be true for its autonomous robots. Since the start button activates a process toward a predetermined final goal, completing the cycle, the autonomous robots of the fabric must operate towards this goal and could manifest their adaptive behaviour only in relation to the achievement of the final output. 
At this point, the argument should be clear: the factory is a program or a system rather than an agent.

Whereas there are no issues with attributing to an AI system the property of \textit{generality}, since numerous reports \cite{bubeck2023sparks}, \cite{openai2023} successfully demonstrate the ability of today’s models to \textit{transfer knowledge} across various disciplines, we should be more sceptical about attributing \textit{agency} to an AI system. 
Ultimately, if the ARC’s experiment did not find any evidence that GPT-4 is capable of replicating itself or becoming difficult to shut down, it is, in the author’s view, because such capabilities are typically exhibited by biological organisms – the only entities that, to date, have demonstrated true agency. This may suggest that, in order to genuinely be capable of acting, an entity must also be a living being. 

\subsection{Implications for the responsibility gap}

If AI cannot be considered an agent, since its operational autonomy does not exhibit any of the characteristics commonly associated with agency in biological organisms, then the significance of the responsibility gap is considerably diminished.
This is because, even though the control relationship between humans and algorithms may appear less direct in the case of machine learning systems, it nevertheless persists.

An argument capable of convincingly closing the gap also for moral responsibility can be explored starting from the problem that the literature faced in the last twenty years, namely, the fact that if one cannot control or foresee the AI outputs, it is then impossible to morally blame him if the AI output produced a harm  \cite{matthias2004responsibility}, \cite{sparrow2007killer}, \cite{bathaee2017artificial}.
Let’s take a famous example from “Killer Robots” \cite{sparrow2007killer}: if a commander activates an autonomous weapon system and later on it kills the wrong person, can we still consider the commander responsible? Based on the teachings of the Start Button Problem, we still can say that the control condition stands, since the commander had a choice whether to activate the autonomous weapon system or not. Simultaneously, by knowing that the system will operate autonomously to accomplish the output after the activation, it is still possible and at the very least foreseeable that, for its autonomy, the autonomous weapon system may make errors in accomplishing the goal.

In this context, the very same case that twenty years ago seemed to create a responsibility gap can today be an object of moral blame. That’s because the control relationship between the human and the machine stands still. It changes its nature since now it is contained in the single act of pressing the start button, however, it is precisely that act that contains the last relevant will which contributed to the production of the harmful event. Without the pressing of the start button, nothing would have happened, and that depends on a human decision.

\subsection{Distinctions between the Start Button Problem and the Chinese Room Argument}

In the end, a short clarification on the distinction between the Start Button Problem and the Chinese Room Argument \cite{searle1980}.
The Chinese Room Argument was intended to challenge a definition of AI at the time qualified as strong AI, which was a hypothesis of a system capable of understanding and passing the Turing Test in its stronger version \cite{turing1950}. The Start Button Problem tackle a different and more recent definition of AI, namely AGI, which refers to a system capable of displaying general intelligence by accomplishing a variety of qualitatively different goals.

In this regard, the Start Button’s finding is that while it is certainly possible to design systems with the property of generality, it is not possible to claim that these systems possess agency property. For this reason, the Start Button Problem, differently from the Chinese Room Argument, is intentionally designed to exclude the presence of an agent within the system.

Unlike the Chinese Room Argument, where a human agent inside the room performs the symbol manipulation operations representing the AI system, the Start Button Problem features only one human agent: the person pressing the start button, positioned outside the factory that represents the AI system. That has been done to demonstrate how a complex system displaying autonomous behaviour is possible despite its lack of agency.

\section{Conclusion}

Having analysed the implications of the Start Button Problem in the previous section, allow us to conclude with a lighter reflection and a few open questions.
The Start Button Problem illustrates the control relationship between humans and AI systems, grounded in the input-output architecture. This architecture requires an external input, whether in the form of activation or goal selection, for any operation to be performed.

When analysed in these terms, human–AI interaction does not appear to generate a discontinuity in traditional models of responsibility ascription. This is because, by pressing the start button, the human assumes full responsibility for the system’s output. The absence of genuine AI agency – revealed through the analysis of the Start Button Problem – remains decisive, even when the system is framed as an instance of artificial general intelligence (AGI).
However, while this analysis closes the traditional responsibility gap framed as a loss of human control over the AI system, it also gives rise to new moral dilemmas. 

Consider a scenario where a person activates a course of action with a very extended causal length. The individual knows that without their initial input nothing would occur and yet chooses to proceed, fully aware that the outcome may be unforeseeable or even harmful. 
We can observe that this one surely is a behaviour suitable for moral blame, since, from the beginning, the person who started the course of action knew that the possibility of unpredictable outcomes exists, and he decided anyway to start it. However, even if he takes full responsibility, how should we evaluate events that manifest months or even years later? While I am inclined to argue that the person remains responsible, it seems clear that this form of responsibility does not easily fit within the traditional legal or moral categories of intent, fault and causation.

This is the new dilemma introduced by the Start Button Problem, made possible by AI algorithms capable of extending the causal consequences of human potential to act across time in unprecedented ways.

\bibliographystyle{abbrvnat}  
\bibliography{references}

\begin{thebibliography}{49}
\providecommand{\natexlab}[1]{#1}
\providecommand{\url}[1]{\texttt{#1}}
\expandafter\ifx\csname urlstyle\endcsname\relax
  \providecommand{\doi}[1]{doi: #1}\else
  \providecommand{\doi}{doi: \begingroup \urlstyle{rm}\Url}\fi

\bibitem[Almada(2024)]{almada2024delegating}
M.~Almada.
\newblock \emph{Delegating the law of artificial intelligence: a procedural account of technology-neutral regulation}.
\newblock PhD thesis, European University Institute, 2024.

\bibitem[Bathaee(2017)]{bathaee2017artificial}
Y.~Bathaee.
\newblock The artificial intelligence black box and the failure of intent and causation.
\newblock \emph{Harv. JL \& Tech.}, 31:\penalty0 889, 2017.

\bibitem[Bender et~al.(2021)Bender, Gebru, McMillan-Major, and Shmitchell]{bender2021dangers}
E.~M. Bender, T.~Gebru, A.~McMillan-Major, and S.~Shmitchell.
\newblock On the dangers of stochastic parrots: Can language models be too big?
\newblock In \emph{Proceedings of the 2021 ACM conference on fairness, accountability, and transparency}, pages 610--623, 2021.
\newblock URL \url{https://doi.org/10.1145/3442188.3445922}.

\bibitem[Biever(2023)]{biever2023chatgpt}
C.~Biever.
\newblock Chatgpt broke the turing test-the race is on for new ways to assess ai.
\newblock \emph{Nature}, 619\penalty0 (7971):\penalty0 686--689, 2023.
\newblock URL \url{https://doi.org/10.1038/d41586-023-02361-7}.

\bibitem[Bubeck et~al.(2023)Bubeck, Chandrasekaran, Eldan, Gehrke, Horvitz, Kamar, Lee, Lee, Li, Lundberg, et~al.]{bubeck2023sparks}
S.~Bubeck, V.~Chandrasekaran, R.~Eldan, J.~Gehrke, E.~Horvitz, E.~Kamar, P.~Lee, Y.~T. Lee, Y.~Li, S.~Lundberg, et~al.
\newblock Sparks of artificial general intelligence: Early experiments with gpt-4.
\newblock \emph{arXiv preprint arXiv:2303.12712}, 2023.
\newblock URL \url{https://doi.org/10.48550/arXiv.2303.12712}.

\bibitem[Buchanan(2024)]{buchanan2024human}
M.~Buchanan.
\newblock Human intelligence is not computable.
\newblock \emph{nature physics}, 20\penalty0 (6):\penalty0 882--882, 2024.

\bibitem[Carneiro(2000)]{carneiro2000transition}
R.~L. Carneiro.
\newblock The transition from quantity to quality: A neglected causal mechanism in accounting for social evolution.
\newblock \emph{Proceedings of the National Academy of Sciences}, 97\penalty0 (23):\penalty0 12926--12931, 2000.
\newblock URL \url{https://doi.org/10.1073/pnas.240462397}.

\bibitem[Chan et~al.(2023)Chan, Salganik, Markelius, Pang, Rajkumar, Krasheninnikov, Langosco, He, Duan, Carroll, et~al.]{chan2023harms}
A.~Chan, R.~Salganik, A.~Markelius, C.~Pang, N.~Rajkumar, D.~Krasheninnikov, L.~Langosco, Z.~He, Y.~Duan, M.~Carroll, et~al.
\newblock Harms from increasingly agentic algorithmic systems.
\newblock In \emph{Proceedings of the 2023 ACM Conference on Fairness, Accountability, and Transparency}, pages 651--666, 2023.
\newblock URL \url{https://doi.org/10.48550/arXiv.2302.10329}.

\bibitem[Chollet(2019)]{chollet2019measure}
F.~Chollet.
\newblock On the measure of intelligence.
\newblock \emph{arXiv preprint arXiv:1911.01547}, 2019.
\newblock URL \url{https://arxiv.org/abs/1911.01547}.

\bibitem[Coeckelbergh(2016)]{coeckelbergh2016responsibility}
M.~Coeckelbergh.
\newblock Responsibility and the moral phenomenology of using self-driving cars.
\newblock \emph{Applied Artificial Intelligence}, 30\penalty0 (8):\penalty0 748--757, 2016.

\bibitem[Coeckelbergh(2020)]{coeckelbergh2020artificial}
M.~Coeckelbergh.
\newblock Artificial intelligence, responsibility attribution, and a relational justification of explainability.
\newblock \emph{Science and engineering ethics}, 26\penalty0 (4):\penalty0 2051--2068, 2020.
\newblock URL \url{https://doi.org/10.1007/s11948-019-00146-8}.

\bibitem[Conradie and Nagel(2024)]{conradie2024no}
N.~H. Conradie and S.~K. Nagel.
\newblock No agent in the machine: being trustworthy and responsible about ai.
\newblock \emph{Philosophy \& Technology}, 37\penalty0 (2):\penalty0 72, 2024.

\bibitem[Dastani and Yazdanpanah(2023)]{dastani2023responsibility}
M.~Dastani and V.~Yazdanpanah.
\newblock Responsibility of ai systems.
\newblock \emph{Ai \& Society}, 38\penalty0 (2):\penalty0 843--852, 2023.

\bibitem[Eilon(1969)]{Elion}
S.~Eilon.
\newblock What is a decision?
\newblock \emph{Management Science}, 16\penalty0 (4):\penalty0 B172--B189, 1969.
\newblock ISSN 00251909, 15265501.
\newblock URL \url{http://www.jstor.org/stable/2628797}.

\bibitem[Floridi(2023)]{floridi2023ai}
L.~Floridi.
\newblock Ai as agency without intelligence: on chatgpt, large language models, and other generative models.
\newblock \emph{Philosophy \& technology}, 36\penalty0 (1):\penalty0 15, 2023.
\newblock URL \url{https://doi.org/10.1007/s13347-023-00621-y}.

\bibitem[Floridi(2025)]{floridi2025ai}
L.~Floridi.
\newblock Ai as agency without intelligence: On artificial intelligence as a new form of artificial agency and the multiple realisability of agency thesis.
\newblock \emph{Philosophy \& Technology}, 38\penalty0 (1):\penalty0 30, 2025.

\bibitem[Franklin and Graesser(1996)]{franklin1996agent}
S.~Franklin and A.~Graesser.
\newblock Is it an agent, or just a program?: A taxonomy for autonomous agents.
\newblock In \emph{International workshop on agent theories, architectures, and languages}, pages 21--35. Springer, 1996.
\newblock URL \url{https://doi.org/10.1007/BFb0013570}.

\bibitem[Goertzel(2014)]{goertzel2014artificial}
B.~Goertzel.
\newblock Artificial general intelligence: concept, state of the art, and future prospects.
\newblock \emph{Journal of Artificial General Intelligence}, 5\penalty0 (1):\penalty0 1, 2014.

\bibitem[Goetze(2022)]{goetze2022mind}
T.~S. Goetze.
\newblock Mind the gap: autonomous systems, the responsibility gap, and moral entanglement.
\newblock In \emph{Proceedings of the 2022 ACM Conference on Fairness, Accountability, and Transparency}, pages 390--400, 2022.

\bibitem[Gon{\c{c}}alves(2023)]{gonccalves2023turing}
B.~Gon{\c{c}}alves.
\newblock The turing test is a thought experiment.
\newblock \emph{Minds and Machines}, 33\penalty0 (1):\penalty0 1--31, 2023.
\newblock URL \url{https://doi.org/10.1007/s11023-022-09616-8}.

\bibitem[Johnson and Verdicchio(2017)]{johnson2017reframing}
D.~G. Johnson and M.~Verdicchio.
\newblock Reframing ai discourse.
\newblock \emph{Minds and Machines}, 27:\penalty0 575--590, 2017.
\newblock URL \url{https://doi.org/10.1007/s11023-017-9417-6}.

\bibitem[Kauffman and Roli(2023)]{kauffman2023consciousness}
S.~A. Kauffman and A.~Roli.
\newblock What is consciousness? artificial intelligence, real intelligence, quantum mind and qualia.
\newblock \emph{Biological Journal of the Linnean Society}, 139\penalty0 (4):\penalty0 530--538, 2023.

\bibitem[K{\"o}nigs(2022)]{konigs2022artificial}
P.~K{\"o}nigs.
\newblock Artificial intelligence and responsibility gaps: What is the problem?
\newblock \emph{Ethics and Information Technology}, 24\penalty0 (3):\penalty0 36, 2022.

\bibitem[Korteling et~al.(2021)Korteling, van~de Boer-Visschedijk, Blankendaal, Boonekamp, and Eikelboom]{korteling2021human}
J.~H. Korteling, G.~C. van~de Boer-Visschedijk, R.~A. Blankendaal, R.~C. Boonekamp, and A.~R. Eikelboom.
\newblock Human-versus artificial intelligence.
\newblock \emph{Frontiers in artificial intelligence}, 4:\penalty0 622364, 2021.
\newblock URL \url{https://doi.org/10.3389/frai.2021.622364}.

\bibitem[LeCun(2022)]{lecun2022path}
Y.~LeCun.
\newblock A path towards autonomous machine intelligence version 0.9. 2, 2022-06-27.
\newblock \emph{Open Review}, 62\penalty0 (1):\penalty0 1--62, 2022.
\newblock URL \url{https://openreview.net/pdf?id=BZ5a1r-kVsf}.

\bibitem[Legg and Hutter(2007)]{legg2007universal}
S.~Legg and M.~Hutter.
\newblock Universal intelligence: A definition of machine intelligence.
\newblock \emph{Minds and machines}, 17:\penalty0 391--444, 2007.
\newblock URL \url{https://doi.org/10.1007/s11023-007-9079-x}.

\bibitem[Matthias(2004)]{matthias2004responsibility}
A.~Matthias.
\newblock The responsibility gap: Ascribing responsibility for the actions of learning automata.
\newblock \emph{Ethics and information technology}, 6:\penalty0 175--183, 2004.
\newblock URL \url{https://doi.org/10.1007/s10676-004-3422-1}.

\bibitem[Mitchell(2024{\natexlab{a}})]{mitchell2024debates}
M.~Mitchell.
\newblock Debates on the nature of artificial general intelligence, 2024{\natexlab{a}}.
\newblock URL \url{https://doi.org/10.1126/science.ado7069}.

\bibitem[Mitchell(2024{\natexlab{b}})]{mitchell2024turing}
M.~Mitchell.
\newblock The turing test and our shifting conceptions of intelligence, 2024{\natexlab{b}}.
\newblock URL \url{https://doi.org/10.1126/science.adq9356}.

\bibitem[Morris et~al.(2024)Morris, Sohl-Dickstein, Fiedel, Warkentin, Dafoe, Faust, Farabet, and Legg]{morris2024}
M.~R. Morris, J.~Sohl-Dickstein, N.~Fiedel, T.~Warkentin, A.~Dafoe, A.~Faust, C.~Farabet, and S.~Legg.
\newblock Position: Levels of agi for operationalizing progress on the path to agi.
\newblock In \emph{Proceedings of the 41st International Conference on Machine Learning}, volume 235 of \emph{PMLR}, pages 36308--36321, 2024.
\newblock URL \url{https://proceedings.mlr.press/v235/morris24b.html}.

\bibitem[Nyholm(2018)]{nyholm2018attributing}
S.~Nyholm.
\newblock Attributing agency to automated systems: Reflections on human--robot collaborations and responsibility-loci.
\newblock \emph{Science and engineering ethics}, 24\penalty0 (4):\penalty0 1201--1219, 2018.

\bibitem[OpenAI(2023)]{openai2023}
OpenAI.
\newblock Gpt-4 technical report.
\newblock \emph{ArXiv preprint}, arXiv:2303.08774, 2023.
\newblock URL \url{https://arxiv.org/abs/2303.08774}.

\bibitem[Roli et~al.(2022)Roli, Jaeger, and Kauffman]{roli2022organisms}
A.~Roli, J.~Jaeger, and S.~A. Kauffman.
\newblock How organisms come to know the world: Fundamental limits on artificial general intelligence.
\newblock \emph{Frontiers in Ecology and Evolution}, 9:\penalty0 806283, 2022.

\bibitem[Russell and Norvig(2020)]{russell2020}
S.~J. Russell and P.~Norvig.
\newblock \emph{Artificial Intelligence: A Modern Approach}.
\newblock Pearson College Div., 4th edition, 2020.

\bibitem[Santoni~de Sio and Mecacci(2021)]{santoni2021four}
F.~Santoni~de Sio and G.~Mecacci.
\newblock Four responsibility gaps with artificial intelligence: Why they matter and how to address them.
\newblock \emph{Philosophy \& technology}, 34\penalty0 (4):\penalty0 1057--1084, 2021.

\bibitem[Sartor et~al.(2016)Sartor, Omicini, et~al.]{sartor2016autonomy}
G.~Sartor, A.~Omicini, et~al.
\newblock The autonomy of technological systems and responsibilities for their use.
\newblock In \emph{Autonomous Weapon Systems. Law, Ethics, Policy}, pages 39--74. Cambridge University Press, 2016.

\bibitem[Searle(1980{\natexlab{a}})]{searle1980}
J.~R. Searle.
\newblock Minds, brains and programs.
\newblock \emph{Behavioral and Brain Sciences}, 3:\penalty0 417--57, 1980{\natexlab{a}}.

\bibitem[Searle(1980{\natexlab{b}})]{searle1980intentionality}
J.~R. Searle.
\newblock The intentionality of intention and action.
\newblock Technical report, University of California, Berkeley, Cognitive Science, 1980{\natexlab{b}}.

\bibitem[Searle(2006)]{searle2006}
J.~R. Searle.
\newblock \emph{Freedom and Neurobiology: Reflections on Free Will, Language, and Political Power}.
\newblock Columbia University Press, 2006.

\bibitem[Shoemaker(2011)]{shoemaker2011attributability}
D.~Shoemaker.
\newblock Attributability, answerability, and accountability: Toward a wider theory of moral responsibility.
\newblock \emph{Ethics}, 121\penalty0 (3):\penalty0 602--632, 2011.

\bibitem[Silver et~al.(2021)Silver, Singh, Precup, and Sutton]{silver2021reward}
D.~Silver, S.~Singh, D.~Precup, and R.~S. Sutton.
\newblock Reward is enough.
\newblock \emph{Artificial intelligence}, 299:\penalty0 103535, 2021.

\bibitem[Solum(2020)]{solum2020legal}
L.~B. Solum.
\newblock Legal personhood for artificial intelligences.
\newblock In \emph{Machine ethics and robot ethics}, pages 415--471. Routledge, 2020.

\bibitem[Sparrow(2007)]{sparrow2007killer}
R.~Sparrow.
\newblock Killer robots.
\newblock \emph{Journal of applied philosophy}, 24\penalty0 (1):\penalty0 62--77, 2007.

\bibitem[Tigard(2021)]{tigard2021there}
D.~W. Tigard.
\newblock There is no techno-responsibility gap.
\newblock \emph{Philosophy \& Technology}, 34\penalty0 (3):\penalty0 589--607, 2021.

\bibitem[Turing(1950)]{turing1950}
A.~M. Turing.
\newblock Computing machinery and intelligence.
\newblock \emph{Mind}, 49:\penalty0 433--460, 1950.

\bibitem[Veneziani(2024)]{veneziani2024}
P.~Veneziani.
\newblock Deleghe di funzioni e culpa in vigilando nella prospettiva della sicurezza del lavoro.
\newblock \emph{CRIMINALIA}, pages 1--28, 2024.
\newblock ISSN 1972-3857.

\bibitem[Vitarelli(2010)]{vitarelli2010}
T.~Vitarelli.
\newblock La disciplina della delega di funzioni.
\newblock In \emph{Il nuovo diritto penale della sicurezza nei luoghi di lavoro}, pages 37--57. Giuffrè, 2010.

\bibitem[Wang et~al.(2024)Wang, Ma, Feng, Zhang, Yang, Zhang, Chen, Tang, Chen, Lin, et~al.]{wang2024}
L.~Wang, C.~Ma, X.~Feng, Z.~Zhang, H.~Yang, J.~Zhang, Z.~Chen, J.~Tang, X.~Chen, Y.~Lin, et~al.
\newblock A survey on large language model based autonomous agents.
\newblock \emph{Frontiers of Computer Science}, 18\penalty0 (6):\penalty0 186345, 2024.
\newblock URL \url{https://doi.org/10.1007/s11704-024-40231-1}.

\bibitem[Wei et~al.(2022)Wei, Tay, Bommasani, Raffel, Zoph, Borgeaud, Yogatama, Bosma, Zhou, Metzler, et~al.]{wei2022}
J.~Wei, Y.~Tay, R.~Bommasani, C.~Raffel, B.~Zoph, S.~Borgeaud, D.~Yogatama, M.~Bosma, D.~Zhou, D.~Metzler, et~al.
\newblock Emergent abilities of large language models.
\newblock \emph{Transactions on Machine Learning Research}, 2022.
\newblock URL \url{https://arxiv.org/abs/2206.07682}.

\end{thebibliography}

\end{document}